\documentclass[conference, a4paper]{IEEEtran}
\IEEEoverridecommandlockouts

\usepackage{cite}
\usepackage{amsmath,amssymb,amsfonts}
\usepackage{algorithmic}
\usepackage{graphicx}
\usepackage{textcomp}
\usepackage{xcolor}
\def\BibTeX{{\rm B\kern-.05em{\sc i\kern-.025em b}\kern-.08em
    T\kern-.1667em\lower.7ex\hbox{E}\kern-.125emX}}
\begin{document}

\title{SR-NeRV: Improving Embedding Efficiency of Neural Video Representation via Super-Resolution}

\author{\IEEEauthorblockN{Taiga Hayami}
\IEEEauthorblockA{\textit{Graduate School of FSE,} \\
\textit{Waseda University}\\
Tokyo, Japan \\
hayatai17@fuji.waseda.jp}
\and
\IEEEauthorblockN{Kakeru Koizumi}
\IEEEauthorblockA{\textit{Graduate School of FSE,} \\
\textit{Waseda University}\\
Tokyo, Japan \\
kkeverio@ruri.waseda.jp}
\and
\IEEEauthorblockN{Hiroshi Watanabe}
\IEEEauthorblockA{\textit{Graduate School of FSE,} \\
\textit{Waseda University}\\
Tokyo, Japan \\
hiroshi.watanabe@waseda.jp}
}
\maketitle

\begin{abstract}
Implicit Neural Representations (INRs) have garnered significant attention for their ability to model complex signals in various domains. 
Recently, INR-based frameworks have shown promise in neural video compression by embedding video content into compact neural networks. 
However, these methods often struggle to reconstruct high-frequency details under stringent constraints on model size, which are critical in practical compression scenarios.
To address this limitation, we propose an INR-based video representation framework that integrates a general-purpose super-resolution (SR) network. 
This design is motivated by the observation that high-frequency components tend to exhibit low temporal redundancy across frames.
By offloading the reconstruction of fine details to a dedicated SR network pre-trained on natural images, the proposed method improves visual fidelity.
Experimental results demonstrate that the proposed method outperforms conventional INR-based baselines in reconstruction quality, while maintaining a comparable model size.
\end{abstract}

\begin{IEEEkeywords}
Implicit neural representation, super-resolution, video representation.
\end{IEEEkeywords}

\section{Introduction}
With the rapid growth of video content, efficient video representation and compression have become critical challenges.
Implicit neural representation (INR) has shown strong potential for modeling complex signals, including 3D shapes and scenes, by representing them as continuous functions through neural networks that map coordinates to signal values.
In the context of video compression, this idea has been extended by embedding video data into neural networks and compressing the network weights. 
Neural Representations for Videos (NeRV) \cite{nerv} introduced a pioneering framework for implicit video representation by learning a frame-wise mapping from temporal indices to video frames.
Owing to its lightweight architecture, NeRV achieves faster decoding compared to conventional video compression techniques.
However, NeRV and subsequent INR-based video representation methods primarily focus on embedding efficiency and often struggle to reconstruct high-frequency components of frames due to the spectral bias characteristic of neural networks \cite{finer}.
Moreover, frame reconstruction via simple upsampling operations often leads to visual distortion.

To address these challenges, we propose an INR-based video representation framework enhanced with a general-purpose super-resolution (SR) model.
The key intuition behind our approach is that high-frequency components such as fine textures and sharp edges exhibit low temporal redundancy, making them difficult to infer reliably from adjacent frames.
To compensate for this, we offload the reconstruction of such details to a dedicated SR model.
This model is pre-trained on a large-scale dataset of natural images and learns to restore plausible high-frequency patterns commonly found in real-world scenes.
To ensure effective integration with the NeRV framework, we apply synthetic degradations, including blur and color distortion, to the training images.
These degradation processes are designed to emulate the characteristic artifacts observed in NeRV reconstructions.
By incorporating the SR model into the NeRV decoding pipeline, our method enhances spatial detail and perceptual quality without increasing the complexity of the core NeRV architecture.
Experimental results demonstrate that the proposed framework outperforms conventional INR-based methods in reconstruction quality while maintaining a comparable model size.

\begin{figure}[tb]
\centerline{\includegraphics[width=1\columnwidth]{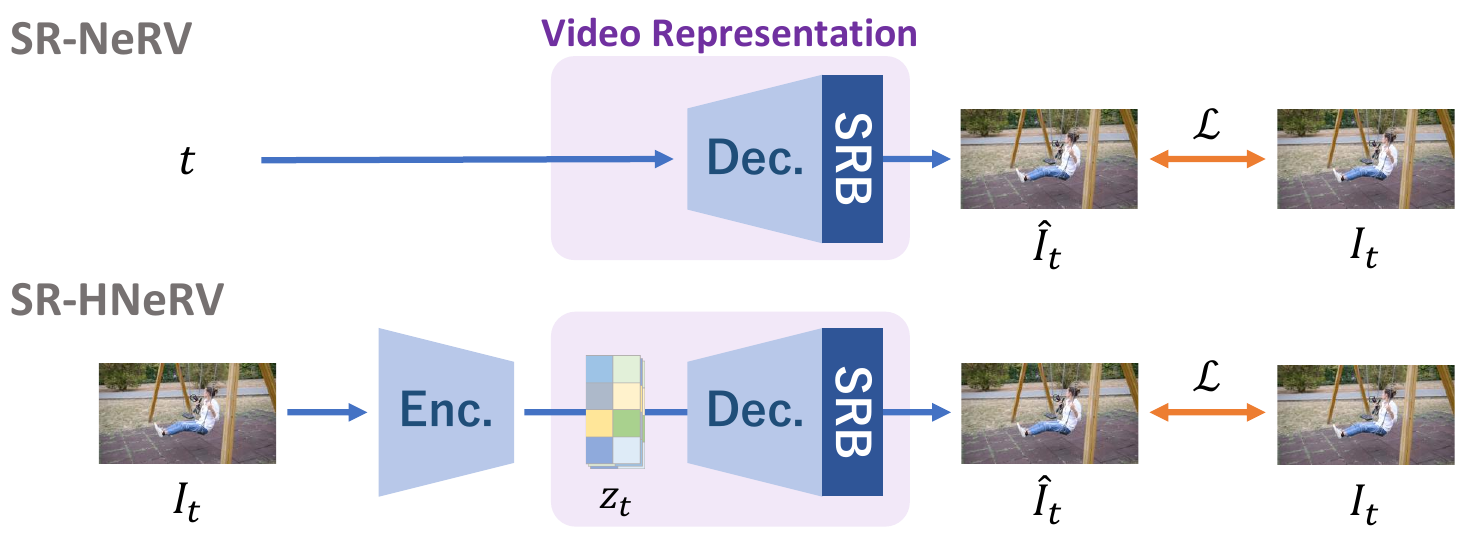}}
\caption{Overview of the proposed SR-enhanced architecture. A Super-Resolution Block (SRB) is integrated into NeRV-based frameworks (NeRV and HNeRV) to improve reconstruction of high-frequency components.}
\label{pipeline}
\end{figure}

\section{Related Works}
\subsection{Neural Representations for Videos}
INRs have been widely adopted for modeling 3D scenes \cite{nerf} and various other signals, due to their coordinate-based formulation that maps spatial or spatiotemporal positions directly to signal values.
In the context of video representation, frame-wise INR methods that map frame indices to entire frames have demonstrated superior training efficiency and faster convergence compared to pixel-wise methods \cite{siren} that learn mappings from coordinates to individual pixel values.
NeRV \cite{nerv} introduced a pioneering framework that employs temporal indices and lightweight neural decoders to generate video frames through implicit representation.
As a subsequent study, HNeRV \cite{hnerv} improves the representational capacity by incorporating video-specific frame features.
Other works explore hierarchical architectures \cite{dsnerv, pnerv} or explicitly separate frequency bands \cite{nvrrrcp, rcnerv} to better capture high-frequency details.
Nevertheless, accurately preserving fine details remains a persistent challenge for NeRV-based approaches.

\subsection{Super-Resolution}
Super-resolution (SR) techniques aim to reconstruct high-resolution images from low-resolution inputs by restoring missing high-frequency components.
Numerous models based on convolutional neural networks (CNNs) and generative adversarial networks (GANs) have been proposed to address this task.
SRCNN \cite{srcnn} is among the earliest CNN-based approaches to demonstrate the effectiveness of end-to-end learning for image upscaling.
To improve computational efficiency, ESPCN \cite{espcn} introduces the sub-pixel convolution (pixel-shuffle) layer, which replaces deconvolution-based upsampling by learning to rearrange low-resolution feature maps into high-resolution outputs.
ESRGAN \cite{esrgan} is a GAN-based method that enhances perceptual quality by employing Residual-in-Residual Dense Blocks (RRDBs), which increase the network’s capacity to model complex textures and fine details.
Given their strength in recovering high-frequency components, SR models offer a promising solution for enhancing NeRV-based video representations, which often struggle to reconstruct fine-grained spatial content.

\section{Proposed Method}

\begin{figure}[tb]
\centerline{\includegraphics[width=0.7\columnwidth]{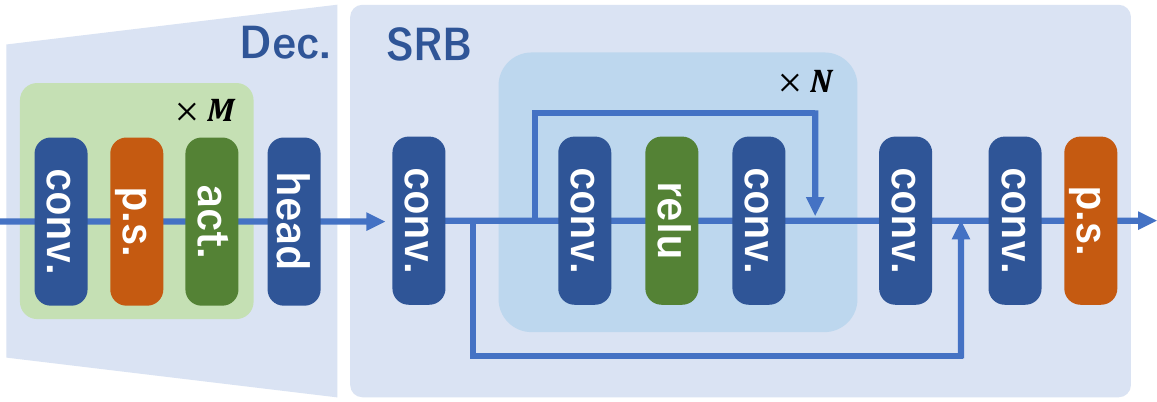}}
\caption{Structure of the decoder and the Super-Resolution Block (SRB). The SRB, which consists of residual blocks and a pixel-shuffle upsampling layer, is applied after the decoder.}
\label{srb}
\end{figure}

We propose a neural video representation method to enhance the quality of reconstructed video frames by integrating a Super-Resolution Block (SRB) into conventional NeRV-based video representation frameworks, including NeRV \cite{nerv} and HNeRV \cite{hnerv}, as illustrated in Fig. \ref{pipeline}.
Our approach addresses two primary limitations of existing NeRV-based methods: the spectral bias of neural network training and the upsampling bottleneck. 
The spectral bias causes the learning process to favor low-frequency components, resulting in poor reconstruction of high-frequency details. 
Furthermore, upsampling from compact latent representations to full-resolution outputs often results in a degradation of visual quality, particularly under tight model size constraints.
To alleviate these issues, we incorporate a general-purpose SR model that reconstructs high-frequency components, which exhibit low temporal redundancy across frames.

The SR model, pre-trained on a generic image dataset, learns to restore plausible high-frequency patterns commonly observed in natural scenes.
To emulate the degraded outputs typically produced by NeRV-based methods, we follow the degradation strategy introduced in NeRFLiX \cite{nerflix} by applying blur and color transformations to low-resolution training inputs.
Formally, given a low-resolution image $I_{LR}$, we generate its degraded version $\tilde{I}_{LR}$ and reconstruct a high-resolution output $\hat{I}_{HR}$ as follows:
\begin{equation}
\tilde{I}_{LR} = \mathcal{C}(\mathcal{B}(I_{LR})),\quad \hat{I}_{HR} = f_{\theta}(\tilde{I}_{LR}),
\end{equation}
where $\mathcal{B}(\cdot)$ denotes a blur operation, and $\mathcal{C}(\cdot)$ represents a color transformation. 
The resulting degraded image $\tilde{I}_{LR}$ is then passed to a neural network $f_{\theta}$, which reconstructs the corresponding high-resolution output $\hat{I}_{HR}$. 
Here, $\theta$ denotes the set of learnable parameters of the model.
As shown in Fig. \ref{pipeline}, our architecture integrates a lightweight SR network, which is structured with residual blocks and pixel-shuffle upsampling to facilitate efficient high-frequency reconstruction.

\begin{table}[tb]
\caption{Comparison of reconstructed frames at \\ different training epochs on the Bunny dataset}
\begin{center}
\begin{tabular}{c|ccccc}
\hline
Epoch    & 300            & 600            & 1200           & 2400           & 4800           \\ \hline\hline
NeRV     & 30.87          & 31.67          & 32.14          & 32.41          & 32.56          \\ 
SR-NeRV  & \textbf{32.66} & \textbf{33.56} & \textbf{34.09} & \textbf{34.64} & \textbf{34.96} \\ \hline
HNeRV    & 35.60          & 36.84          & 37.47          & 37.85          & 38.15          \\ 
SR-HNeRV & \textbf{36.66} & \textbf{37.85} & \textbf{38.17} & \textbf{38.53} & \textbf{38.82} \\ \hline
\end{tabular}
\end{center}
\label{psnr-epoch}
\end{table}

\begin{figure*}[tb]
\centerline{\includegraphics[width=2.0\columnwidth]{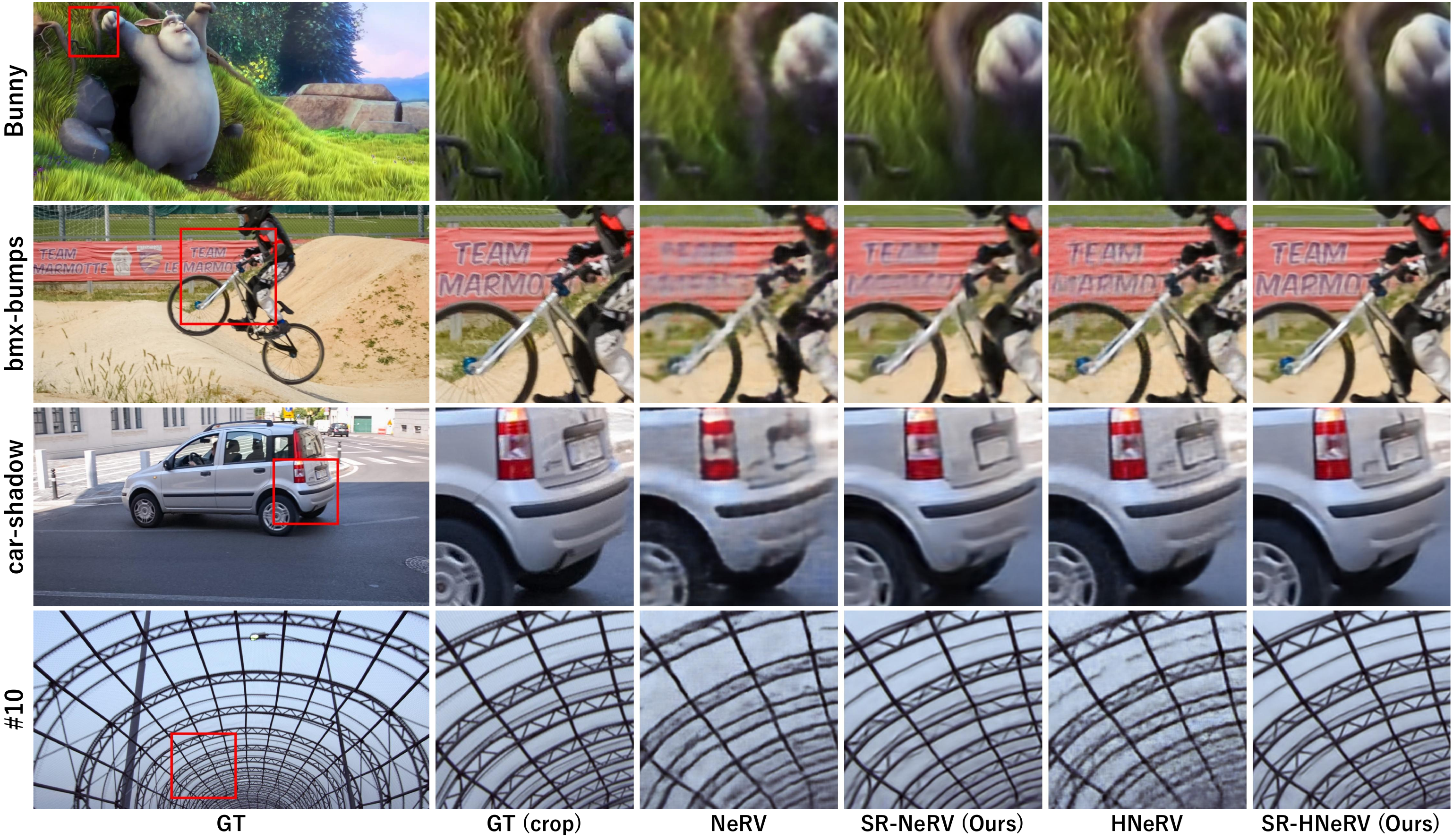}}
\caption{Qualitative comparison of reconstructed frames on Bunny, DAVIS, and MCL-JCV datasets. SR-NeRV and SR-HNeRV produce sharper outputs with fewer artifacts compared to NeRV and HNeRV.}
\label{result}
\end{figure*}

The trained SR model is appended after the decoder of the NeRV-based model, as formulated below:
\begin{equation}
\hat{I}_{t} = f_\theta(g_\phi(z_t)),
\end{equation}
where $z_t$ denotes the frame embedding at time step $t$, and $\hat{I}_t$ is the reconstructed output frame.
In this formulation, $g_\phi$ represents the NeRV decoder parameterized by $\phi$, and $f_\theta$ denotes the SR model with parameters $\theta$.
To ensure stable training and improve embedding efficiency, the SR model remains frozen during the first half of the training process.
In the second half, we fine-tune the model for better adaptation to the specific characteristics of the target video content.

\section{Experiment}
In this section, we describe the training setups for both the SR model and the NeRV-based video representation model.
We then evaluate the proposed method on multiple video datasets using both quantitative and qualitative metrics.

\subsection{Setup}
The SR model is trained on the DIV2K dataset \cite{div2k}, a widely used benchmark for image super-resolution.
During training, the SR model learns to reconstruct high-resolution images with a $2\times$ spatial scale from a low-resolution input degraded by blur and color transformations.
The reconstruction is supervised using the corresponding ground-truth image, and the model is optimized by minimizing the L1 loss between the predicted and ground-truth images.

\begin{table}[tb]
\caption{Quantitative comparison of reconstruction quality \\ on the DAVIS dataset}
\begin{center}
\begin{tabular}{c|cccc}
\hline
Method   & Model Size [M]   & PSNR $\uparrow$   & MS-SSIM $\uparrow$ & LPIPS $\downarrow$ \\ \hline\hline
NeRV     & 1.51 (\ \ -\ \ ) & 28.55             & 0.8814             & 0.3740             \\ 
SR-NeRV  & 1.49 (0.16)      & \textbf{29.26}    & \textbf{0.8924}    & \textbf{0.3616}    \\ \hline
HNeRV    & 1.50 (\ \ -\ \ ) & 30.91             & 0.9198             & 0.2999             \\ 
SR-HNeRV & 1.50 (0.16)      & \textbf{31.97}    & \textbf{0.9348}    & \textbf{0.2863}    \\ \hline
\end{tabular}
\end{center}
\label{davis}
\end{table}

For video representation, we adopt two NeRV-based methods, NeRV \cite{nerv} and HNeRV \cite{hnerv}, as baseline models.
We integrate the proposed Super-Resolution Block (SRB) into these models to construct SR-NeRV and SR-HNeRV, respectively.
Experiments are conducted on three datasets: Big Buck Bunny (Bunny), DAVIS \cite{davis}, and MCL-JCV \cite{mcl}.
The Bunny dataset is a single 132-frame animated video sequence. 
The DAVIS dataset comprises 50 video sequences (25-104 frames each) of natural scenes, while the MCL-JCV dataset contains 30 sequences (120-150 frames each) from video codec evaluation content. 
All frames are center-cropped to a resolution of $640 \times 1280$ for training.
To ensure fair comparison, we adopt the same architectural configurations as the original NeRV and HNeRV models, modifying the stride settings when integrating the SRB to maintain consistent output resolution.
Specifically, the stride configurations are set as follows: $[5, 4, 2, 2]$ for NeRV, $[5, 4, 4, 2, 2]$ for HNeRV, $[5, 2, 2, 2]$ for SR-NeRV, and $[5, 4, 2, 2, 2]$ for SR-HNeRV.
Each video sequence is trained and evaluated independently to assess the model’s ability to represent individual videos.
Model optimization is performed using the L2 loss, and all other hyperparameters follow the standard settings from prior NeRV implementations \cite{hnerv}.

\begin{table}[tb]
\caption{Quantitative comparison of reconstruction quality \\ on the MCL-JCV dataset}
\begin{center}
\begin{tabular}{c|cccc}
\hline
Method   & Model Size [M]   & PSNR $\uparrow$   & MS-SSIM $\uparrow$ & LPIPS $\downarrow$ \\ \hline\hline
NeRV     & 1.51 (\ \ -\ \ ) & 31.69             & 0.9249             & 0.3648             \\ 
SR-NeRV  & 1.50 (0.16)      & \textbf{32.45}    & \textbf{0.9349}    & \textbf{0.3478}    \\ \hline
HNeRV    & 1.51 (\ \ -\ \ ) & 33.68             & 0.9458             & 0.3096             \\ 
SR-HNeRV & 1.48 (0.16)      & \textbf{34.62}    & \textbf{0.9559}    & \textbf{0.2968}    \\ \hline
\end{tabular}
\end{center}
\label{mcl}
\end{table}

\subsection{Video Representation}
The reconstruction quality of videos on the Bunny dataset at various training epochs is illustrated in Table \ref{psnr-epoch}.
The integration of the proposed SRB into the baseline NeRV model leads to a substantial improvement in frame quality.
The reconstruction results for the DAVIS and MCL-JCV datasets are shown in Tables \ref{davis} and \ref{mcl}, respectively.
Each table reports the average PSNR, MS-SSIM, and LPIPS across all video sequences, excluding those where training failed to converge. 
The total number of model parameters is 1.5 million, with the contribution of the SRB component noted in parentheses.
Across all datasets and evaluation metrics, the proposed SR-NeRV and SR-HNeRV consistently outperform their respective baselines, NeRV and HNeRV. 
These results confirm that incorporating the SRB module significantly enhances video reconstruction fidelity, with negligible impact on model complexity.

We further evaluate the visual quality of reconstructed frames through qualitative comparisons.
Fig. \ref{result}  presents example frames from the Bunny, DAVIS, and MCL-JCV datasets, illustrating the ground-truth frames, the outputs from baseline NeRV/HNeRV models, and the results of our SR-enhanced models.
Our method consistently recovers finer details and produces noticeably sharper frames compared to the baselines.
For example, in the DAVIS sequence ``bmx-bumps'', SR-HNeRV resolves text in the background that HNeRV fails to read. 
In the ``car-shadow'' sequence from DAVIS, our model eliminates blurring artifacts present in the NeRV reconstruction.
Similarly, in sequence \#10 of the MCL-JCV dataset, both SR-NeRV and SR-HNeRV more effectively preserve edge sharpness than their baseline counterparts.
These qualitative and quantitative results demonstrate that the proposed SR-augmented models can more accurately reconstruct high-frequency details in video content.

\subsection{Ablation Study}
We performed an ablation study to investigate the effect of the fine-tuning schedule for the SRB.
Specifically, we varied the epoch at which fine-tuning of the SRB begins (i.e., the epoch at which the SRB is ``unfrozen'' during training) and evaluated the final reconstruction quality on the Bunny dataset.
Table~\ref{bunny} reports the resulting PSNR values for different fine-tuning start points.
A fine-tuning start epoch of 0 indicates that the SRB is trained from the beginning (i.e., no freezing), whereas a value of 300 corresponds to the case where the SRB remains frozen throughout training (i.e., no fine-tuning).
We observe that starting fine-tuning at epoch 50 achieves the highest PSNR, suggesting that the timing of SRB fine-tuning plays a crucial role in reconstruction quality.
We hypothesize that initiating fine-tuning around epoch 50 allows the NeRV decoder to first converge to a stable reconstruction, after which the SRB can more effectively refine the output by performing content-specific corrections.
In contrast, delaying fine-tuning for too long may cause a mismatch between the NeRV decoder’s output and the SRB’s expected input (i.e., a simulated low-quality frame), thereby limiting the overall reconstruction performance.

\section{Conclusion}
In this paper, we propose a video representation framework that incorporates a super-resolution model to enhance the reconstruction of high-frequency components in video frames. 
By incorporating an SR model pre-trained on natural images, which implicitly captures high-frequency patterns, into a conventional NeRV-based architecture, we achieve a more accurate representation of fine details.
Furthermore, we introduced a degradation simulation strategy during SR training to mitigate artifacts specific to NeRV-based methods.
Experimental results demonstrate that our approach improves the quality of reconstructed frames across a variety of video sequences.
For future work, we plan to develop more effective SR architectures and integration strategies specifically optimized for NeRV-based frameworks.

\begin{table}[tb]
\caption{PSNR comparison on the Bunny dataset under different fine-tuning start epochs for SR-NeRV and SR-HNeRV}
\begin{center}
\begin{tabular}{c|ccccc}
\hline
Fine-tuning Start Epoch & 0     & 50             & 100   & 150   & 300   \\ \hline\hline
SR-NeRV                 & 31.64 & \textbf{32.66} & 31.97 & 31.36 & 29.74 \\ \hline
SR-HNeRV                & 36.45 & \textbf{36.66} & 36.32 & 36.01 & 34.33 \\ \hline
\end{tabular}
\end{center}
\label{bunny}
\end{table}

\vspace{12pt}
\end{document}